# Photo-thermionic effect in vertical graphene heterostructures

M. Massicotte[1], P. Schmidt[1,*], F. Vialla[1,*], K. Watanabe[2], T. Taniguchi[2], K.J. Tielrooij[1], F.H.L. Koppens[1,3]

[1] ICFO-Institut de Ciencies Fotoniques, The Barcelona Institute of Science and Technology, 08860 Castelldefels (Barcelona), Spain.

[2] National Institute for Materials Science, 1-1 Namiki, Tsukuba 305-0044, Japan.

[3] ICREA – Institució Catalana de Recerça i Estudis Avancats, Barcelona, Spain.

* These authors contributed equally

**Finding alternative optoelectronic mechanisms that overcome the limitations of conventional semiconductor devices is paramount for detecting and harvesting low-energy photons. A highly promising approach is to drive a current from the thermal energy added to the free-electron bath as a result of light absorption. Successful implementation of this strategy requires a broadband absorber where carriers interact among themselves more strongly than with phonons, as well as energy-selective contacts to extract the excess electronic heat. Here we show that graphene-WSe$_2$-graphene heterostructure devices offer this possibility through the photo-thermionic effect: the absorbed photon energy in graphene is efficiently transferred to the electron bath, leading to a thermalized hot carrier distribution. Carriers with energy higher than the Schottky barrier between graphene and WSe$_2$ can be emitted over the barrier, thus creating photocurrent. We experimentally demonstrate that the photo-thermionic effect enables detection of sub-bandgap photons, while being size-scalable, electrically tunable, broadband and ultrafast.**

# Introduction

Since the discovery of the photoelectric effect in the late 19th century[1], a great number of photodetectors that rely on the emission of photoexcited charge carriers have been proposed. These carriers – sometimes referred to as hot carriers although they are not thermalized with the electron bath – are typically injected over a Schottky barrier between a metal and a semiconductor, allowing detection of photons with energy lower than the semiconductor bandgap (see Fig. 1a). This process, called internal photoemission, has led to the development of visible and near-infrared photodetectors[2,3], which have recently been combined with plasmonic enhancement schemes[4–8]. However, the efficiency of this mechanism drops for photon energy lower than the Schottky barrier height $\Phi_B$ (ref. [9]) and is limited by the ability to extract the carriers before they lose their initial energy, which in metals typically occurs on a timescale of approximately 100 fs (ref. [10]).

A promising way to overcome these limitations is to make use of the excess thermal energy contained in the electron bath. This energy arises from the thermalization of photoexcited carriers with other carriers, which results in a hot carrier distribution with a well-defined temperature $T_e$. For increasing $T_e$, a larger fraction of carriers can overcome the Schottky barrier, creating a current via thermionic emission (Fig. 1b). In this scheme, even photons with energies below $\Phi_B$ can lead to an increase in $T_e$ and subsequently to carrier emission. However, in order to reach high $T_e$, the hot carriers must be weakly coupled to the surrounding phonon bath[11].

Graphene, which has recently emerged as an excellent platform for converting photons into hot carriers[12], has the ideal properties to implement this scheme. Graphene presents strong electron-electron interactions leading to carrier thermalization within less than 50 fs (ref. [13,14]), where a large fraction (larger than 50%) of the initial energy of photoexcited carriers is transferred to the electronic system[15]. This efficient carrier heating creates a thermalized hot carrier state that is relatively long-lived (longer than 1 ps)[16], owing to weak coupling to the lattice and the environment. These thermalized carriers can thus reach temperatures significantly higher than the phonon bath temperature ($T_e > T_{ph}$) even under continuous-wave (CW) excitation[17] (see Supplementary Note 1 and Supplementary Figure 1). Moreover, the tunability of the graphene Fermi energy gives control over the height of the Schottky barrier. For these reasons, graphene was recently proposed as a promising material for efficient and tunable thermionic emission of hot carriers[18–20].

Here, we use graphene/WSe$_2$/graphene van der Waals heterostructures to detect low-energy photons (with a wavelength up to 1.5 µm) through photo-thermionic (PTI) emission. Figure 1b shows in detail how the PTI photocurrent is generated: Photons are absorbed by graphene, creating electron-hole pairs, which then rapidly equilibrate into a thermalized carrier distribution with an elevated electron temperature $T_e$ compared to the temperature of the lattice $T_{ph}$ and the environment $T_0$; Carriers within this distribution with an energy larger than the Schottky barrier height $\Phi_B$ at the graphene/WSe$_2$ interface can be injected into the WSe$_2$ and travel to the other graphene layer. The number of carriers with sufficient energy scales with $e^{\frac{-\phi_B}{k_B T_e}}$, where $k_B$ is the Boltzmann constant.

## Results

### Device structure

In our device, WSe$_2$ – a transition metal dichalcogenide with a band gap $E_g \sim 1.3$ eV – provides an energy barrier between the two graphene sheets with low interfacial defects and reduced Fermi-level pinning. The active device (depicted schematically in Fig. 1c) is encapsulated between layers of hexagonal boron nitride (hBN) which provides a clean, charge-free environment for the graphene and WSe$_2$ flakes[21]. The device is equipped with an electrostatic bottom gate ($V_G$) that enables control of the Fermi energy $\mu$ and thereby $\Phi_B$ of (mainly) the bottom graphene. All measurements presented in the main text are obtained from one particular device comprising a 28-nm-thick WSe$_2$ flake (see Fig. 1d) and are performed at room temperature with a quasi-CW laser source, unless otherwise mentioned (see Methods). We have studied devices with WSe$_2$ flakes of various thicknesses ($L$ = 2.2 to 40 nm) and obtained similar results, consistent with the PTI effect (see Supplementary Note 2 and Supplementary Figures 2 and 3).

**Photocurrent measurements**

The PTI process is driven by the light-induced increase of the thermal energy of the electron gas ($k_B T_e$). Signatures of this mechanism are readily visible in the data presented in Fig. 2. First, the photocurrent spectrum of Fig. 2a shows a sizable, spectrally flat response for photon energies well below the band gap of WSe$_2$ ($E_{photon}$ < $E_g$). That is expected from a thermally driven photocurrent, given the uniform absorption of graphene in the visible, near-infrared range and the fact that $k_B T_e$ is independent of $E_{photon}$ for constant power[15,22]. Furthermore, the photocurrent generated in this sub-bandgap regime exhibits a striking superlinear dependence on laser power $P$ (Fig. 2b-c). This is a direct consequence of the thermal activation of carriers over the Schottky barrier, which, in first approximation, scales exponentially with $P$ (see Methods). In contrast, the photocurrent in the above-bandgap regime ($E_{photon}$ > $E_g$) varies strongly with $E_{photon}$ and scales linearly with $P$. This photoresponse is characteristic of light absorption in WSe$_2$ and transfer of photoexcited carriers to the graphene electrodes, a process driven by the potential drop across the WSe$_2$ layer[23–25].

Alternative photocurrent generation mechanisms are less likely to contribute to the observed sub-bandgap photocurrent. To verify this, we measured a device with a Au/WSe$_2$ interface, where photocurrent is generated by internal photoemission of non-thermalized photoexcited carriers (see Supplementary Note 3 and Supplementary Figure 4). This device shows a strong dependence on $E_{photon}$ along with a cut-off energy at $E_{photon}$ = $\Phi_B$, and a linear power dependence – clearly at odds with our observations for G/WSe$_2$/G (where G stands for graphene) devices. We note that multi-photon absorption followed by charge transfer could also lead to a superlinear power dependence, but the laser intensity required to induce significant two-photon absorption in either graphene or WSe$_2$ is at least 1-2 orders of magnitude higher than the one used in our experiment (smaller than 1 GWcm$^{-2}$) (ref. [26,27]). Similarly, the photo-thermoelectric and bolometric effects could generate sub-bandgap photocurrent, but both would have a sublinear – rather than the observed superlinear – power dependence[16,28].

To further verify that the sub-bandgap photocurrent stems from the PTI effect, we perform time-resolved photocurrent measurements by varying the time delay $\Delta t$ between two sub-picosecond laser pulses generated by a Ti:Sapphire laser (see Supplementary Note 4 and Supplementary Figure 5). From the dynamics of the positive correlation signal (due to the superlinear power dependence) in Fig. 2d, we extract a characteristic decay time $\tau_{\text{cool}}$ of 1.3 ps, which is on the order of the cooling time of hot carriers in graphene[16,22]. All together the observations presented in Fig. 2 suggest that the sub-bandgap, superlinear, picosecond photocurrent is governed by the PTI effect.

**Electrical tuning of the PTI effect**

In contrast to bulk metal-semiconductor systems, this graphene-based heterostructure offers the possibility to tune the Schottky barrier, and therefore the magnitude of the PTI photocurrent, using the interlayer bias voltage ($V_B$) and gate voltage ($V_G$). Applying these voltages is necessary in order to generate a finite photocurrent, as it breaks the symmetry of our device which is composed of two G/WSe$_2$ Schottky barriers with opposite polarity. As the IR photocurrent maps ($E_{\text{photon}}$ = 0.8 eV) in Fig. 3a and b indicate, the interlayer voltage $V_B$ essentially controls over which of the two Schottky barriers hot carriers are injected: for $V_B$ = -0.6 V (Fig. 3a), the photoactive region corresponds to the area where the top graphene overlaps with the WSe$_2$ layer (G$_T$/WSe$_2$), whereas the interface with the bottom graphene (G$_B$/WSe$_2$) is photoactive for $V_B$ = +0.6 V (Fig. 3b). In Fig. 3c-d we examine the photocurrent originating from regions containing a single G/WSe$_2$ interface, thus allowing us to assess each Schottky barrier individually. To create a current, hot carriers need to be emitted over the G/WSe$_2$ interface and subsequently transported along the WSe$_2$ layer and collected by the other graphene electrode, as illustrated in the insets of Fig. 3c-d. When the interlayer bias $V_B$ makes this process energetically favorable, each Schottky barrier gives rise to a photocurrent with a specific sign. The photocurrent generated in the G/WSe$_2$/G region (Fig. 3e) exhibits both signs as it stems from charge injection over both top and bottom Schottky barriers. From the photocurrent sign associated with each layer, we deduce that hot electrons, rather than holes, are predominantly emitted over both Schottky barriers. This is expected given the work functions of graphene and electron affinity of WSe$_2$ (ref. [29]).

One of the hallmarks of thermionic emission is its exponential dependence on the Schottky barrier height. In our device, the gate voltage $V_G$ provides a crucial way of enhancing the photocurrent by controlling the height of the G$_B$/WSe$_2$ Schottky barrier via the tuning of the Fermi energy of G$_B$. As Fig. 3f demonstrates, doping the bottom graphene layer with electrons by applying a positive gate voltage $V_G$ effectively lowers $\Phi_B$ and results in a strong increase in photocurrent. At high $V_G$ (low $\Phi_B$), the device reaches a responsivity of up to 0.12 mAW$^{-1}$ at wavelength $\lambda$ = 1500 nm, which, for 0.5% light absorption in graphene[30], translates into an internal quantum efficiency (*IQE*) of 2%. These figures of merit are similar to those obtained in devices using the in-plane photo-thermoelectric effect[31] and can be further improved by adjusting the relevant physical parameters as discussed below.

**Theoretical model of the PTI effect**
The gate tunability of the PTI process and its distinct power dependence allow for a quantitative comparison of our measurements with a Schottky barrier model based on the Landauer transport formalism[32] (see Methods). In this model, the photocurrent depends on the fraction of carriers with enough energy to overcome the barrier, governed by $T_e$ and $\Phi_B$, and on the carrier injection time $\tau_{\text{inj}}$. The values for $\Phi_B(V_G)$ are extracted from temperature-dependent dark current measurements (Fig. 4a) and are consistent with a band offset $\Phi_0$ of 0.54 eV (ref. [29]). For simplicity, we assume that heat in the electronic system dissipates through a single, rate-limiting cooling pathway characterized by a thermal conductance $\Gamma$, such that under steady-state conditions the increase in temperature is proportional to $P/\Gamma$ (see Supplementary Note 1).

Figure 4b compares the measured and fitted *PC* as a function of $\Phi_B(V_G)$ and laser power. This two-dimensional fit yields a carrier injection time $\tau_{inj}$ = 47 ± 10 ps and a thermal conductance $\Gamma$ = 0.5 ± 0.3 MWm$^{-2}$K$^{-1}$. This value of $\tau_{inj}$ is almost identical to the one found for ideal G/Si Schottky barriers[32], while the one obtained for $\Gamma$ matches the predicted thermal conductance of G/hBN interfaces due to electron coupling with SPP phonons[33] and is also consistent with disorder-enhanced supercollisions with acoustic phonons[16] (see Supplementary Note 1). The excellent agreement between model and experiment is clearly visible in Fig. 4c-d. We note that the same measurements were performed at other ambient temperatures ($T_0$ = 230 and 330 K) and the analysis yields very similar results (see Supplementary Note 5 and Supplementary Figure 6).

## Discussion

The device modeling and extracted physical parameters provide important insights into how to improve the efficiency of the PTI process. They also explain why this mechanism dominates the photoresponse of graphene/semiconductor heterostructures, while being absent for metal/semiconductor devices. The reason is that the thermal conductance $\Gamma$ of our graphene-based device is more than 2 orders of magnitude smaller than the thermal conductance due to electron-phonon coupling in thin (approximately 10 nm) metal films[10] (see Supplementary Note 3). Hence, thermalized hot carriers in metals do not reach a sufficiently high temperature to generate significant PTI photocurrent. Strategies to substantially increase the device efficiency include further reduction of the thermal conductance in graphene-based devices, for example, by using a non-polar encapsulating material[33]. Likewise, the efficiency of the process can be readily improved by lowering $\Phi_B$. Indeed, we find that the PTI efficiency increases by one order of magnitude (up to 20 %) by extrapolating the *IQE* to higher $T_e$ (approximately 1000 K) or lower $\Phi_B$ (approximately 0.06 eV, see Supplementary Note 6 and Supplementary Figure 7). Moreover, our model suggests that the efficiency can be greatly enhanced by reducing the carrier injection time $\tau_{inj}$, which is related to the coupling energy between adjacent layers. The long $\tau_{inj}$ obtained from our fit appears to be one of the main factors limiting the observed *IQE* and is presumably due to momentum mismatch between electronic states in the two adjacent materials. The interlayer transfer of charge carriers and heat in van der Waals heterostructures is currently not well understood and further studies are needed in order to unveil the limits of the PTI efficiency.

We finally note that the PTI effect shows some similarities to photon-enhanced thermionic emission (PETE), with the important differences that for PETE the photoexcited carriers are in thermal equilibrium with the lattice of a hot semiconductor and are emitted over a vacuum energy barrier[34,35]. There are also important resemblances between the PTI mechanism and the concept of hot-carrier solar cells, since both require decoupling of the electron and phonon baths and energy-selective contacts[11,36]. Both PETE devices and hot-solar cells have an interesting potential for power conversion, but harvesting low-energy photons is limited by the bandgap of the semiconducting absorber. Interestingly, in our PTI device, which has a very simple geometry and operates at room temperature, we also find a gate-dependent open-circuit voltage (of the photocurrent) of up to 0.17 V with a fill factor of 38%. This effect, observable in Fig 3e, opens up a promising avenue for infrared energy harvesting using graphene as the active material[37]. Furthermore, the PTI mechanism should work over an extremely broad wavelength range, including the mid-infrared and far-infrared (terahertz) regions and can be used for ultrafast photodetection, given that the signal recovery time is on the order of picoseconds. Finally, these vertical thermionic devices have a scalable active area and can be easily integrated with conventional and flexible solid-state devices. These features make the photo-thermionic effect a highly promising mechanism for a plethora of optoelectronic applications[38].

## Methods

### Device fabrication and optoelectronics measurements

The heterostructures are fabricated the same way as described in ref. 24. Photocurrent is generated by focusing a supercontinuum laser (NKT Photonics SuperK extreme, repetition rate $f$ = 40 MHz and pulse duration $dt$ = 100 ps) with a microscope objective (Olympus LCPlanN 50x) on the device. The photocurrent is measured using a preamplifier and a lock-in amplifier synchronized with a mechanical chopper at 117 Hz.

### PTI emission model

In the reverse-bias regime, the Schottky barrier model based on the Landauer transport formalism[32] (see Supplementary Note 7) predicts that the current density $J$ thermionically emitted over a Schottky barrier of height $\Phi_B$ at temperature $T$ is

$$J(T) = \frac{2}{\pi} \frac{e_0}{\tau_{\text{inj}}} \left(\frac{k_B T}{\hbar v_F}\right)^2 \left(\frac{\Phi_0}{k_B T} + 1\right) exp\left(\frac{-\Phi_B}{k_B T}\right) \quad (1)$$

where $e_0$ is the elementary charge, $k_B$ is the Boltzmann's constant, $\hbar$ is the reduced Planck's constant, $v_F$ is the graphene Fermi velocity, $\Phi_0$ is the band offset at the G/WSe$_2$ interface (0.54 eV) and $\tau_{\text{inj}}$ is the charge injection time. In our experiment, the thermionic photocurrent ($PC$) we measure is produced by the increase of electronic temperature $\Delta T = T_e - T_0$ upon illumination of the device with a quasi-CW laser at $\lambda$ = 1500 nm. Hence, it follows that the photocurrent is $PC = AD[J(T_0 + \Delta T) - J(T_0)]$, where $A$ is the area of the laser beam (laser spot size of 1.75 μm), $D$ is the duty cycle ($D = dt \cdot f = 0.04\,\%$) and $T_0$ is the ambient temperature. Using these equations and assuming $T_0 \gg \Delta T$, one can show that $PC \propto \Delta T + \frac{\phi_B}{2k_B T_0^2} \Delta T^2 + ...$, which makes evident the superlinear behavior of the photocurrent. Finally, we assume that the rise in electronic temperature created by each pulse is $\Delta T = \alpha \eta_{\text{heat}} P / AD\Gamma$, where $\alpha$ is the light absorption in graphene (0.5%), $\eta_{\text{heat}}$ is the fraction of absorbed energy that is transferred to the electron bath (approximately 70%)[15], $P$ is the average power of the laser and $\Gamma$ is the thermal conductance of the rate-limiting thermal dissipation step (see Supplementary Note 1).

**Data availability**

The data that support the findings of this study are available from the corresponding author upon request.

**References**

1. Hertz, H. Ueber einen Einfluss des ultravioletten Lichtes auf die electrische Entladung. *Ann. Phys. Chem* **267,** 983–1000 (1887).

2. Peters, D. W. An infrared detector utilizing internal photoemission. *Proc. IEEE* **55,** 704–705 (1967).

3. Scales, C. & Berini, P. Thin-film Schottky barrier photodetector models. *Quantum Electron. IEEE J.* **46,** 633–643 (2010).

4. Brongersma, M. L., Halas, N. J. & Nordlander, P. Plasmon-induced hot carrier science and technology. *Nature Nanotech.* **10,** 25–34 (2015).

5. Clavero, C. Plasmon-induced hot-electron generation at nanoparticle/metal-oxide interfaces for photovoltaic and photocatalytic devices. *Nature Photon.* **8,** 95–103 (2014).

6. Goykhman, I., Desiatov, B., Khurgin, J., Shappir, J. & Levy, U. Locally Oxidized Silicon Surface-Plasmon Schottky Detector for Telecom Regime. *Nano Lett.* **11,** 2219–2224 (2011).

7. Knight, M. W., Sobhani, H., Nordlander, P. & Halas, N. J. Photodetection with active optical antennas. *Science* **332,** 702–704 (2011).

8. Goykhman, I. *et al.* On-Chip Integrated, Silicon–Graphene Plasmonic Schottky Photodetector with High Responsivity and

Avalanche Photogain. *Nano Lett.* **16,** 3005–3013 (2016).

9. Fowler, R. H. The Analysis of Photoelectric Sensitivity Curves for Clean Metals at Various Temperatures. *Phys. Rev.* **38,** 45–56 (1931).

10. Qiu, T. Q. & Tien, C. L. Heat Transfer Mechanisms During Short-Pulse Laser Heating of Metals. *J. Heat Transfer* **115,** 835–841 (1993).

11. Ross, R. T. & Nozik, A. J. Efficiency of hot-carrier solar energy converters. *J. Appl. Phys.* **53,** 3813–3818 (1982).

12. Voisin, C. & Plaçais, B. Hot carriers in graphene. *J. Phys. Condens. Matter* **27,** 160301 (2015).

13. Breusing, M. *et al.* Ultrafast nonequilibrium carrier dynamics in a single graphene layer. *Phys. Rev. B* **83,** 153410 (2011).

14. Brida, D. *et al.* Ultrafast collinear scattering and carrier multiplication in graphene. *Nature Commun.* **4,** 1987 (2013).

15. Tielrooij, K. J. *et al.* Photoexcitation cascade and multiple hot-carrier generation in graphene. *Nature Phys.* **9,** 248–252 (2013).

16. Graham, M. W., Shi, S.-F., Ralph, D. C., Park, J. & McEuen, P. L. Photocurrent measurements of supercollision cooling in graphene. *Nature Phys.* **9,** 103–108 (2012).

17. Gabor, N. M. *et al.* Hot carrier-assisted intrinsic photoresponse in graphene. *Science* **334,** 648–52 (2011).

18. Liang, S.-J. & Ang, L. K. Electron Thermionic Emission from

Graphene and a Thermionic Energy Converter. *Phys. Rev. Appl.* **3,** 1–8 (2015).

19. Rodriguez-Nieva, J. F., Dresselhaus, M. S. & Levitov, L. S. Thermionic Emission and Negative d*I*/d*V* in Photoactive Graphene Heterostructures. *Nano Lett.* **15,** 1451–1456 (2015).

20. Rodriguez-Nieva, J. F., Dresselhaus, M. S. & Song, J. C. W. Hot-carrier convection in graphene Schottky junctions. Preprint at http://arxiv.org/abs/1504.0721 (2015)

21. Dean, C. R. *et al.* Boron nitride substrates for high-quality graphene electronics. *Nature Nanotech.* **5,** 722–726 (2010).

22. Tielrooij, K. J. *et al.* Generation of photovoltage in graphene on a femtosecond timescale through efficient carrier heating. *Nature Nanotech.* **10,** 437–443 (2015).

23. Britnell, L. *et al.* Strong Light-Matter Interactions in Heterostructures of Atomically Thin Films. *Science.* **340,** 1311–1314 (2013).

24. Yu, W. J. *et al.* Highly efficient gate-tunable photocurrent generation in vertical heterostructures of layered materials. *Nature Nanotech.* **8,** 952–958 (2013).

25. Massicotte, M. *et al.* Picosecond photoresponse in van der Waals heterostructures. *Nature Nanotech.* **11,** 42–46 (2015).

26. Chen, W., Wang, Y. & Ji, W. Two-Photon Absorption in Graphene Enhanced by the Excitonic Fano Resonance. *J. Phys. Chem. C* **119,** 16954–16961 (2015).

27. Zhang, S. *et al.* Direct Observation of Degenerate Two-Photon Absorption and Its Saturation in WS2 and MoS2 Monolayer and Few-Layer Films. *ACS Nano* **9,** 7142–7150 (2015).

28. Yan, J. *et al.* Dual-gated bilayer graphene hot-electron bolometer. *Nature Nanotech.* **7,** 472–478 (2012).

29. Kim, K. *et al.* Band Alignment in WSe2–Graphene Heterostructures. *ACS Nano* **9,** 4527–4532 (2015).

30. Stauber, T., Peres, N. M. R. & Geim, A. K. Optical conductivity of graphene in the visible region of the spectrum. *Phys. Rev. B* **78,** 085432 (2008).

31. Freitag, M., Low, T. & Avouris, P. Increased responsivity of suspended graphene photodetectors. *Nano Lett.* **13,** 1644–1648 (2013).

32. Sinha, D. & Lee, J. U. Ideal Graphene/Silicon Schottky Junction Diodes. *Nano Lett.* **14,** 4660–4664 (2014).

33. Low, T., Perebeinos, V., Kim, R., Freitag, M. & Avouris, P. Cooling of photoexcited carriers in graphene by internal and substrate phonons. *Phys. Rev. B* **86,** 045413 (2012).

34. Schwede, J. W. *et al.* Photon-enhanced thermionic emission for solar concentrator systems. *Nature Mater.* **9,** 762–767 (2010).

35. Schwede, J. W. *et al.* Photon-enhanced thermionic emission from heterostructures with low interface recombination. *Nature Commun.* **4,** 1576 (2013).

36. Würfel, P. Solar energy conversion with hot electrons from impact

ionisation. *Sol. Energy Mater. Sol. Cells* **46,** 43–52 (1997).

37. Nelson, C. a, Monahan, N. R. & Zhu, X.-Y. Exceeding the Shockley–Queisser limit in solar energy conversion. *Energy Environ. Sci.* **6,** 3508 (2013).

38. Koppens, F. H. L. *et al.* Photodetectors based on graphene, other two-dimensional materials and hybrid systems. *Nature Nanotech.* **9,** 780–793 (2014).

**Acknowledgements**
The authors are grateful to Qiong Ma, Pablo Jarillo-Herrero, Mark Lundeberg and Ilya Goykhman for valuable discussions. M.M. thanks the Natural Sciences and Engineering Research Council of Canada (PGSD3-426325-2012). P.S. acknowledges financial support by a scholarship from the 'la Caixa' Banking Foundation. F.V. acknowledges financial support from Marie-Curie International Fellowship COFUND and ICFOnest program. K.T. acknowledges financial support from Mineco (FIS2014-59639-JIN). F.K. acknowledges support by Fundacio Cellex Barcelona, the ERC Career integration grant (294056, GRANOP), the ERC starting grant (307806, CarbonLight), the Mineco grants RYC-2012-12281 and FIS2013-47161-P, and support by the EC under the Graphene Flagship (contract no. CNECT-ICT-604391).

**Author contributions**
M.M. and F.H.L.K. conceived and designed the experiments. M.M., P.S. and F.V. fabricated the samples, carried out the experiments and M.M. performed the data analysis. K.W. and T.T provided boron nitride crystals. M.M., F.V., K.J.T., P.S. and F.H.L.K discussed the results and co-wrote the manuscript.

**Competing financial interests**
The authors declare no competing financial interests.

**Figure captions**
*Figure 1* **The photo-thermionic effect and device structure a)** Simplified band diagram illustrating the internal photoemission process taking place at a metal-semiconductor interface. Non-thermalized photoexcited carriers in metal with sufficient energy to overcome the Schottky barrier $\Phi_B$ can be injected into the semiconductor before they lose their initial energy (within 100 fs for conventional metals[10]). The portion of the energy band filled by electrons and the bandgap of the semiconductor are shaded in blue and pale orange, respectively. Low (high) energy photon and the electronic transition following their absorption are represented by red (green) sinusoidal and vertical arrows. The out-of-equilibrium electron distributions $n(E)$ resulting from these processes are illustrated on the left hand side with the corresponding colours. Photoexcited electrons are depicted by blue dots and their possible transfer path is represented by blue dashed arrows. **b)** Simplified band diagram of the PTI effect at a G/WSe$_2$ interface. The ultrafast thermalization of photoexcited carriers in graphene gives rise to a hot-electron distribution $n(E)$ with a lifetime longer than 1 ps. As the number of electrons in the hot tail (yellow shaded area) of $n(E)$ increases, more electrons are emitted over the Schottky barrier $\Phi_B$, which generates a larger thermionic current (represented by the horizontal arrow). The colour gradient from blue to yellow illustrates the heat contained in the electron distribution. The offset between the graphene neutrality point and WSe$_2$ conduction edge is denoted by $\Phi_0$ and was experimentally determined to be 0.54 eV (ref. 28). In both **a** and **b**, **c)** Schematic of the heterostructure on a 285-nm-thick SiO$_2$/Si substrate, to which a gate voltage ($V_G$) is applied to modify the Fermi level $\mu$ of the bottom graphene. An interlayer bias voltage ($V_B$) between the top (G$_T$) and bottom (G$_B$) graphene flakes can be applied, and current or photocurrent flowing through G$_B$ is measured. **d)** Optical image of a heterostructure composed of a 28-nm-thick WSe$_2$ flake. The top and bottom hBN flakes are 10 and 70 nm thick, respectively. For clarity, graphene flakes are shaded in grey and outlined by a black dashed line, whereas WSe$_2$ is coloured in orange and outlined by an orange line. The scale bar is 5 µm.

*Figure 2* **Experimental signatures of photo-thermionic emission a)** Photocurrent (*PC*) spectrum measured at room temperature in the G/WSe$_2$/G region with laser power $P$ = 90 µW, $V_B$ = 0.6 V and $V_G$ = 0 V (same conditions for **b** and **c**). The insets illustrate the absorption process taking place in the different photoresponse regimes: absorption in WSe$_2$ for $E_{photon} > E_g$ and absorption in graphene for $E_{photon} < E_g$. The transition between these two regimes is represented by the background color gradient, where red (blue) corresponds to the graphene (WSe$_2$) absorption regimes. The vertical orange dashed line corresponds to the energy of the bulk WSe$_2$ bandgap. **b)** Power dependence of the photocurrent for various values of photon energy $E_{photon}$. The dots represent experimental data and the solid lines are power law fits ($PC \propto P^\alpha$) obtained with a fit range $P$ = 70 to 120 µW . **c)** Fitted power law index $\alpha$ vs. photon energy, showing the transition from linear to superlinear power dependence. This transition occurs around $E_{photon} = E_g$, the indirect bandgap of WSe$_2$. The error bars correspond to the standard deviation obtained from the linear fit. **d)** Time-resolved photocurrent change $\Delta PC(\Delta t) = PC(\Delta t) - PC(\Delta t \rightarrow \infty)$, measured using the setup and technique described in ref. 24 with an average laser power of 260 µW (wavelength 800 nm), at low temperature (30 K) and bias ($V_B$ = 0.04 V) in order to suppress the contribution of the photocurrent originating from WSe$_2$ absorption. Experimental data are represented by orange dots and the solid black line is a decaying exponential fit with time constant $\tau_{cool}$ =1.3 ps. Inset: Same data and fit in logarithmic scale.

*Figure 3* **Tunable photo-thermionic response a,b)** *PC* maps of the device shown in Figure 1d measured with an interlayer bias voltage $V_B$ of (**a**) -0.6 V and (**b**) 0.6 V, and $V_G$ = 0 V. The graphene flakes are outlined by black dotted lines and the WSe$_2$ flake by solid orange lines, as in Figure 1d. The scale bars are 3 µm. **c,d,e)** *PC* vs. $V_B$ and $V_G$ measured on single Schottky barriers formed by (**c**) top or (**d**) bottom graphene and WSe$_2$, as well as (**e**) double G/WSe$_2$/G interfaces. The coloured circle (green, red and blue) in the upper right corner of each measurement corresponds to the position of the focused laser beam which are indicated on the *PC* maps **a** and **b**. The black dashed line in **e** indicates where *PC* is null. All measurements are scaled to the same colour bar. Insets of **c** and **d**: Side view of the heterostructure illustrating the generation and transport of hot carriers from one graphene flake to the other. Insets of **e**: Band diagrams depicting the PTI effect in G/WSe$_2$/G for $V_B$ < 0 (left) and $V_B$ > 0 (right). **f)** *PC* vs. $V_G$ taken from **e** at $V_B$ = 0.6 V. Inset: Band diagrams of the G$_B$/WSe$_2$ Schottky barrier at low (bottom) and high (top) $V_G$ illustrating the increase of PTI emission resulting from the lowering of $\Phi_B$. All measurements are performed at room temperature, with $E_{photon}$ = 0.8 eV and *P* = 110 µW.

*Figure 4* **Comparison between data and photo-thermionic model a)** Schottky barrier height $\Phi_B$ vs. $V_G$ extracted from the temperature-dependence of dark current measurements. Inset: Arrhenius plot of the dark current at different $V_G$ and $V_B$ = 0.36 V. Experimental data are represented by blue dots and the solid black lines are linear fits. The error bars in the main panel correspond to the standard deviation obtained from these fits. **b)** *PC* vs. $\Phi_B$ and laser power *P*, measured (left plot) and according to our PTI model (right plot). *PC* is measured at room temperature, with $E_{photon}$ = 0.8 eV and $V_B$ = 0.36 V. **c)** *PC* vs. $\Phi_B$ for different values of *P* and **d)** *PC* vs. *P* for various values of $\Phi_B$ taken from **b**. The data points correspond to the experiment and the solid black lines to the model. The upper horizontal axis shows the rise in electronic temperature $\Delta T = T_e - T_0$ (extracted from the fit of the model to the experiment). **d.** Insets of **c** and **d**: Same experimental data and theoretical curves in logarithmic scale.

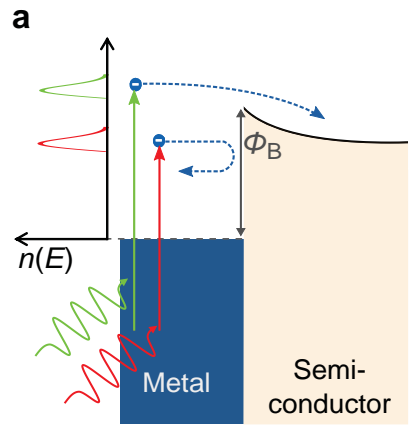
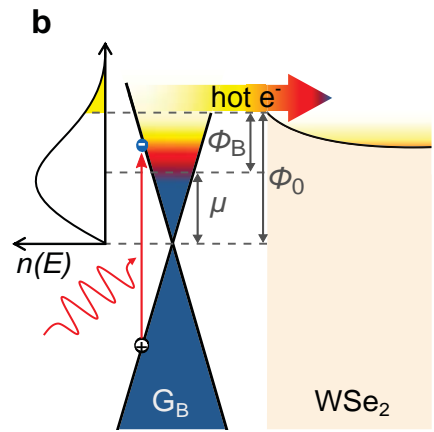
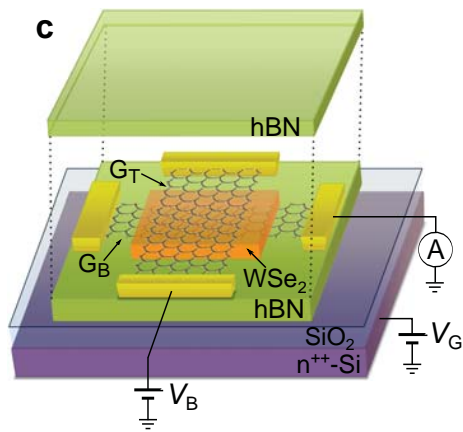
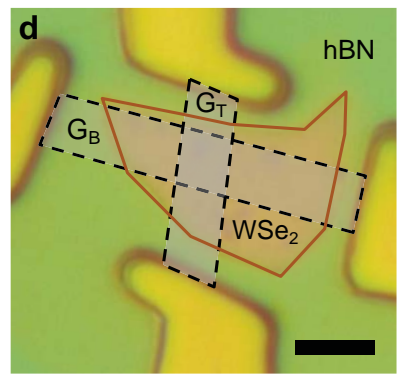

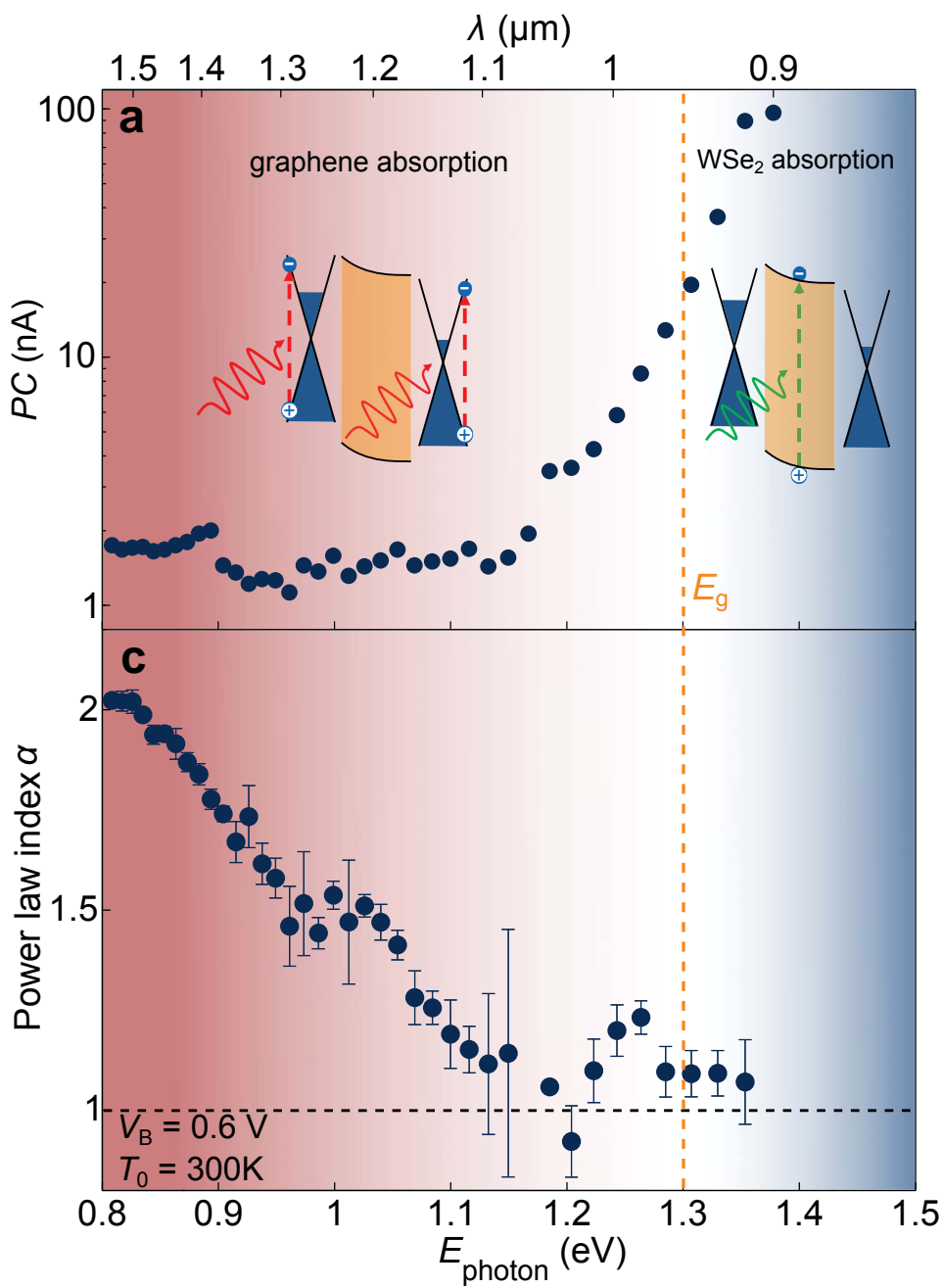
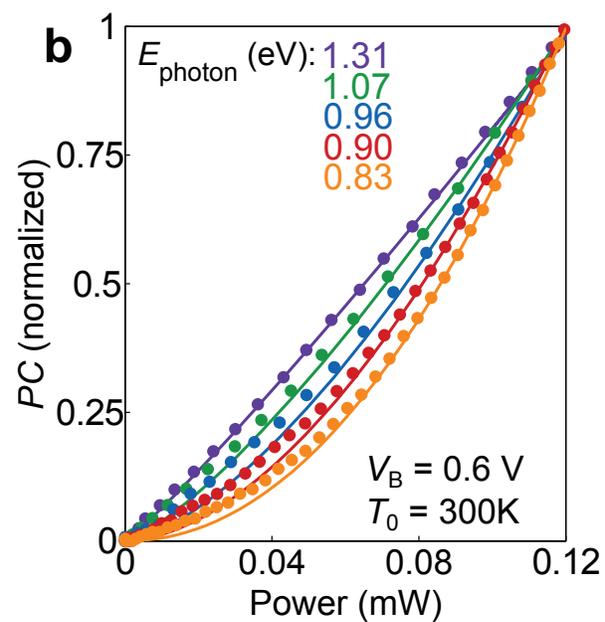
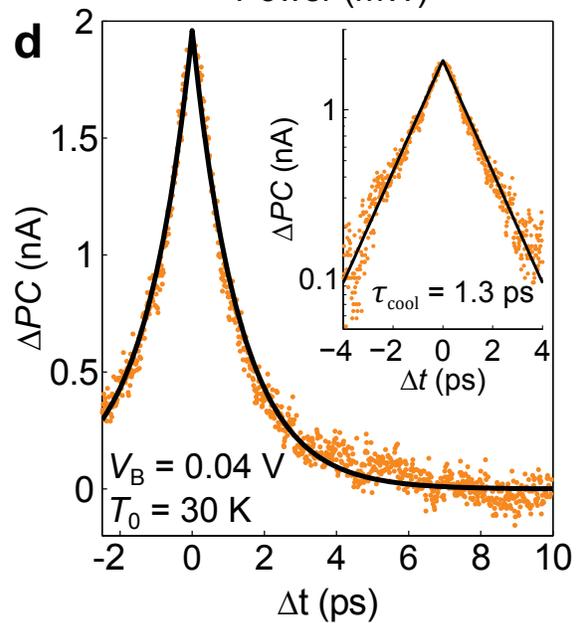

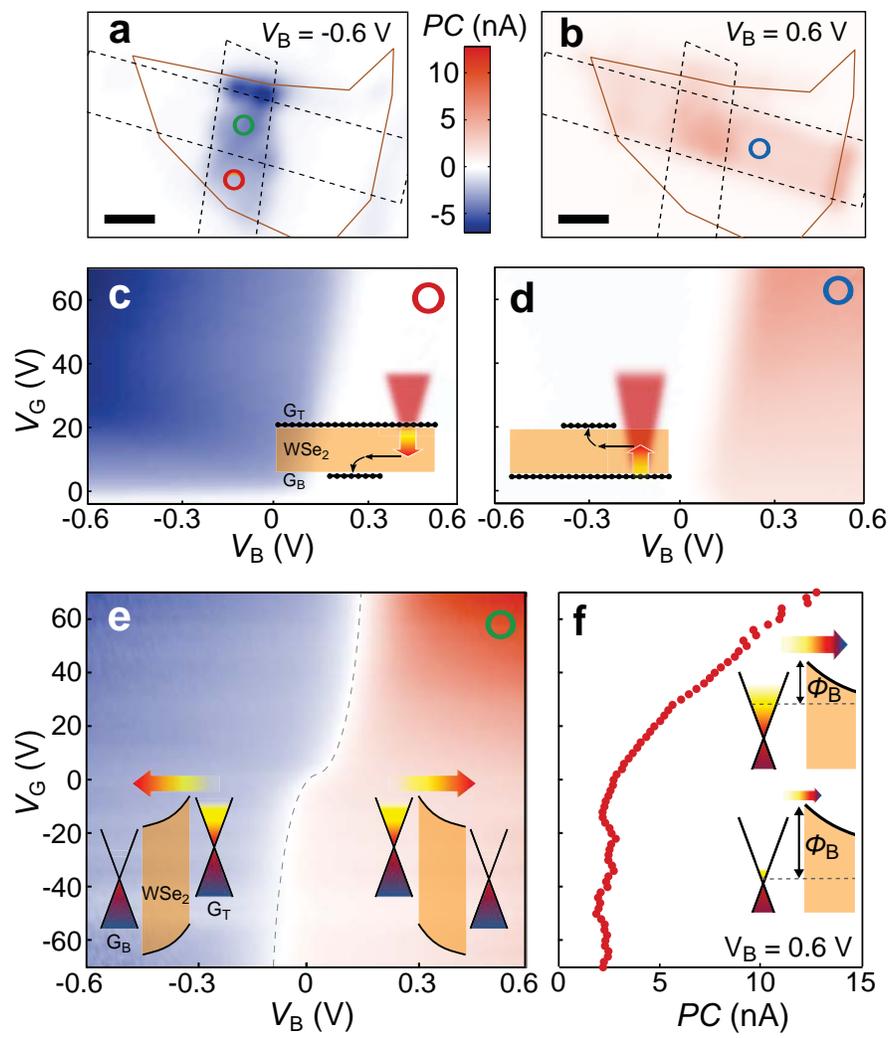

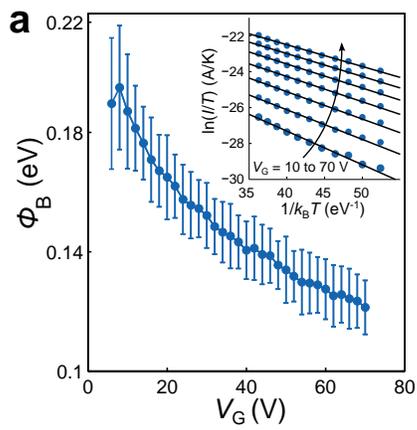
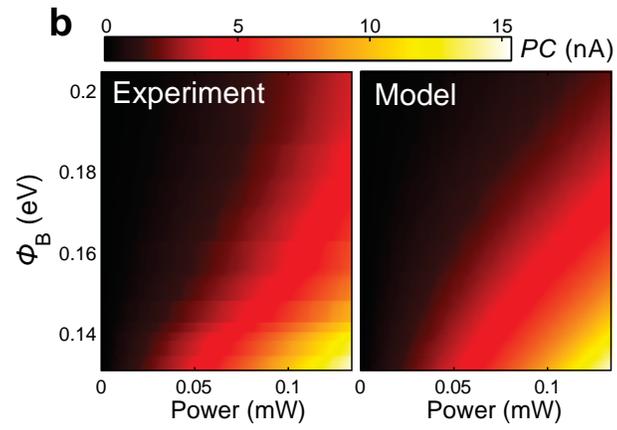
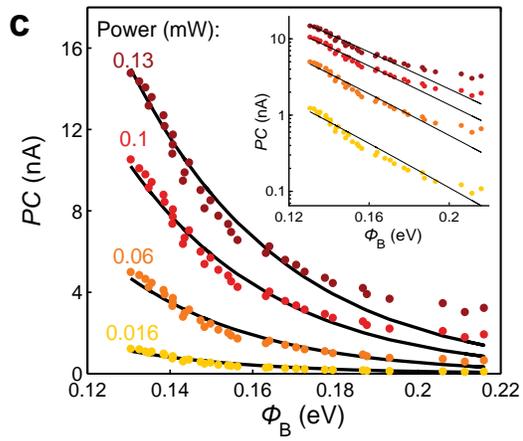
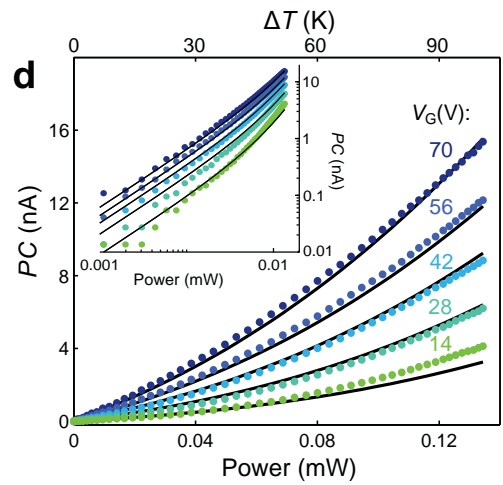

# Supplementary Figures

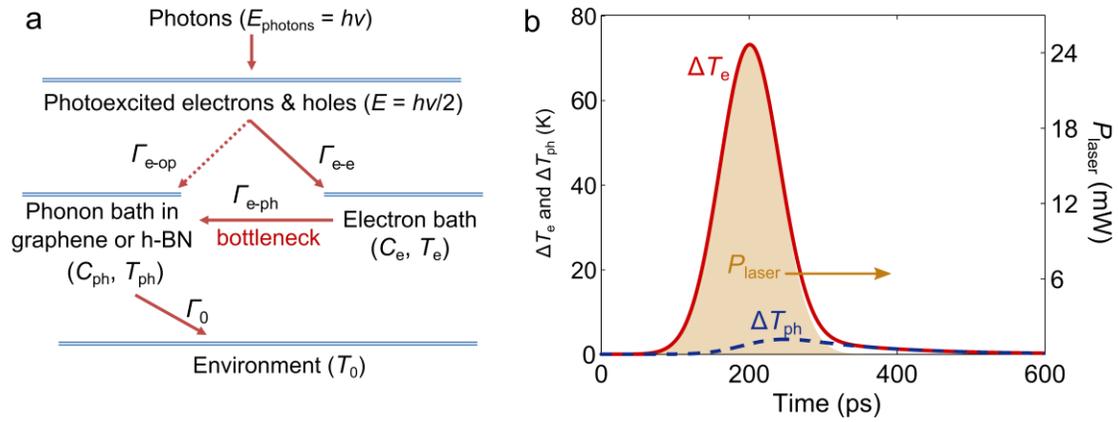

**Supplementary Figure 1: Heating and cooling pathways of hot carriers in graphene. a)** Schematics illustrating the cooling pathway of photoexcited carriers. Detailed descriptions are provided in the text of Supplementary Note 1. **b)** Time dependence of the rise in electron ($\Delta T_e = T_e - T_0$, red solid line) and phonon ($\Delta T_{ph} = T_{ph} - T_0$, blue dotted line) temperature calculated with the model illustrated in **a** under a quasi-CW pulse (full width at half maximum (FWHM) duration $dt = 100$ ps, average laser power $P = 100$ μW and repetition rate $f = 40$ MHz) at $T_0 = 300$K, graphene Fermi level $\mu = 0.2$ eV. The pale yellow area represents the instantaneous laser power $P_{laser}$ of a single pulse centered at 200 ps.

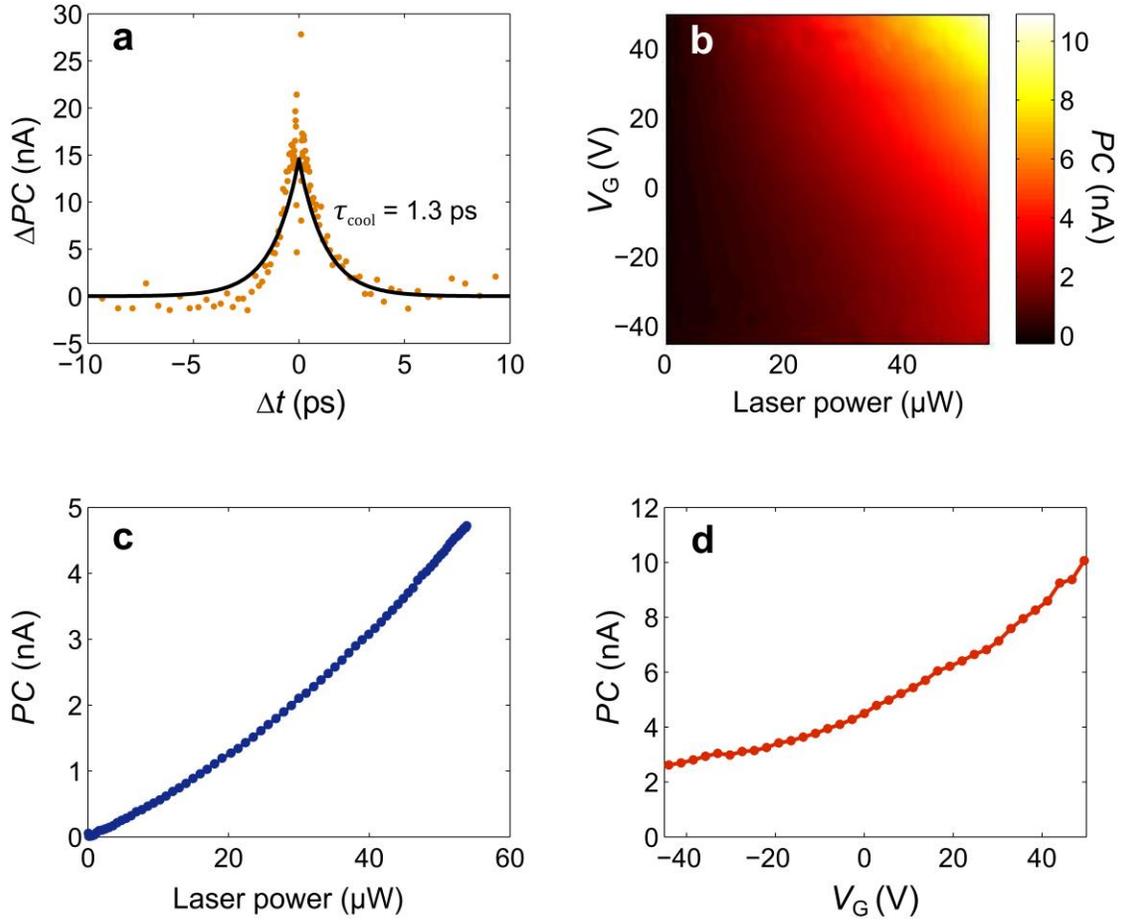

**Supplementary Figure 2: PTI photocurrent in G/2.2-nm-thick WSe$_2$/G heterostructure. a)** Photocurrent autocorrelation measurement performed with an average laser power $P$ = 600 µW at $T_0$ = 30 K and bias voltage $V_B$ = 0.5 V. The decay of $\Delta PC = PC(\Delta t) - PC(\Delta t \rightarrow \infty)$ is fitted with an exponential with time constant $\tau_{cool}$ =1.3 ± 0.1 ps (black solid line). **b)** *PC* vs. laser power and gate voltage $V_G$ measured at bias voltage $V_B$ = 0.7 V and $T_0$ = 30 K, with laser wavelength $\lambda$ = 1300 nm. **c)** *PC* vs. laser power for $V_G$ = 0 V and (**d**) *PC* vs. $V_G$ for $P$ = 50 µW taken from (**b**).

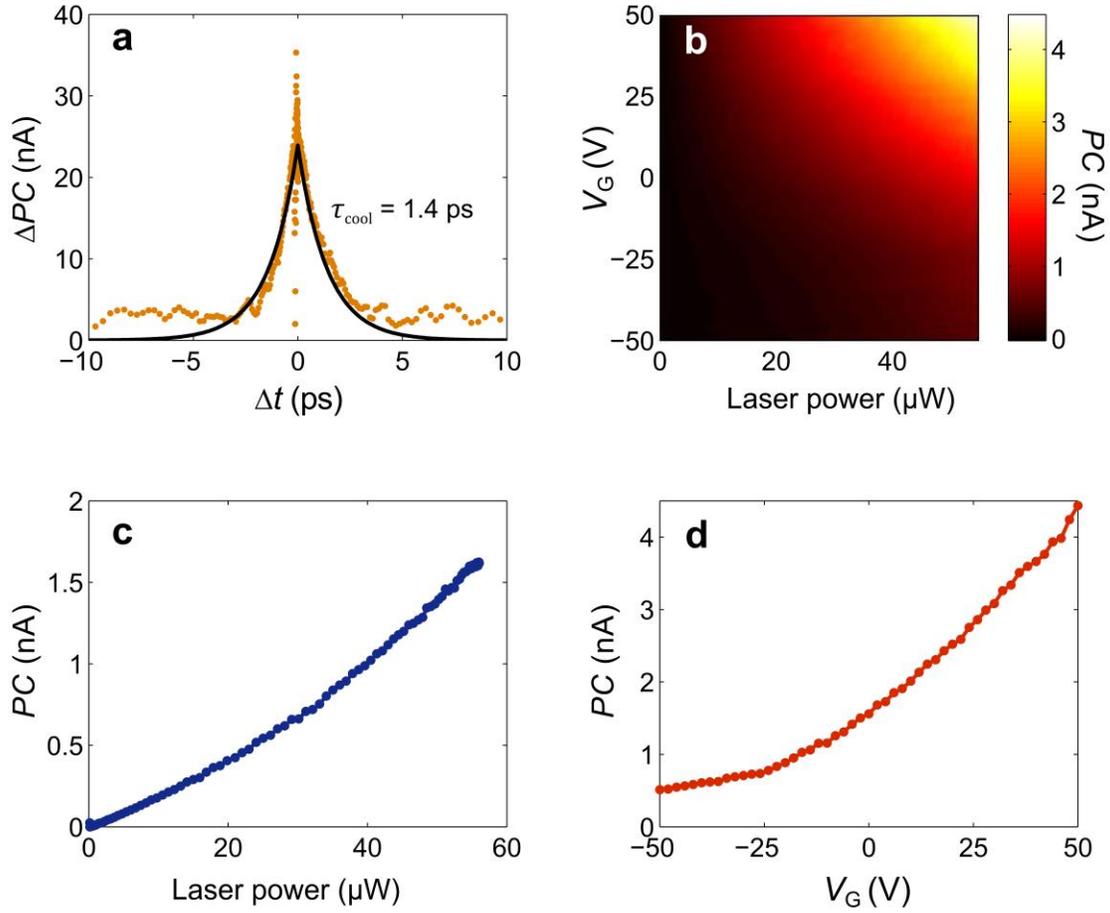

**Supplementary Figure 3: PTI photocurrent in G/7.4-nm-thick WSe2/G heterostructure. a)** Photocurrent autocorrelation measurement performed with an average laser power $P$ = 825 μW at $T_0$ = 300 K and bias voltage $V_B$ = 0.06 V. The decay of $\Delta PC = PC(\Delta t) - PC(\Delta t \rightarrow \infty)$ is fitted with an exponential with time constant $\tau_{cool}$ =1.4 ± 0.1 ps (black solid line). **b)** $PC$ vs. laser power and gate voltage $V_G$ measured at $V_B$ = 0.5 V and $T_0$ = 35 K, with laser wavelength $\lambda$ = 1300 nm. **c)** $PC$ vs. laser power for $V_G$ = 0 V and (**d**) $PC$ vs. $V_G$ for $P$ = 50 μW taken from (**b**).

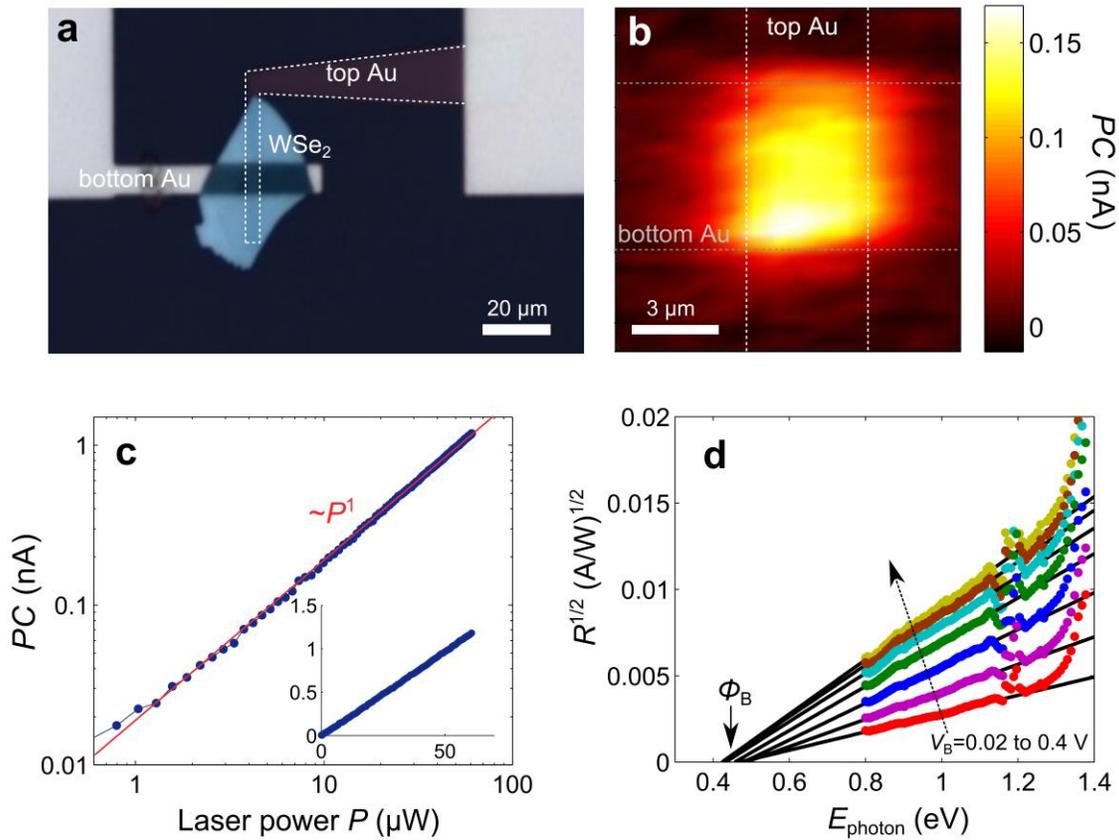

**Supplementary Figure 4: Photocurrent measurements in Au/20-nm-thick WSe$_2$/Au heterostructure. a)** Optical image of the device. **b)** Photocurrent map performed at $T_0$ = 300 K and $V_B$ = 0.2 V, with a laser wavelength $\lambda$ = 1500 nm and power $P$ = 10 μW. The position of the top and bottom Au electrodes is indicated by the white and gray dotted lines, respectively. **c)** Log-log plot of $PC$ vs. laser power measured at $T_0$ = 300 K and $V_B$ = 0.2 V, with $\lambda$ = 1500 nm. The red solid line corresponds to a linear power dependence. Inset: same data on linear scale. **d)** Square root of the responsivity $R$ vs. photon energy $E_{photon}$ at bias voltage $V_B$ from 0.02 V (red) to 0.4 V (yellow). The black solid lines are linear fits to the data.

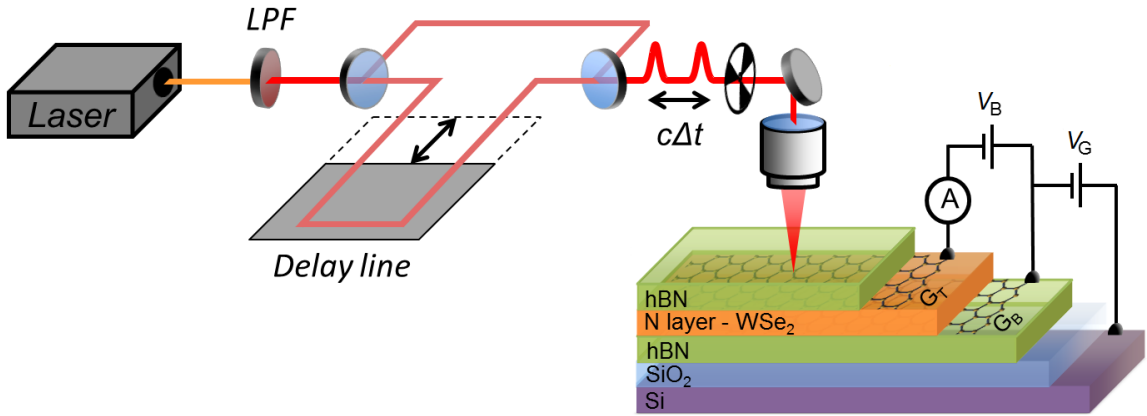

**Supplementary Figure 5: Time-resolved photocurrent measurement setup.** A Ti:sapphire laser generates ultrashort and broadband pulses which can be spatially delayed using a motorized delay stage by a distance $c\Delta t$ where $c$ is the speed of light. A 800-nm long pass filter (LPF) is inserted to improve the PTI signal.

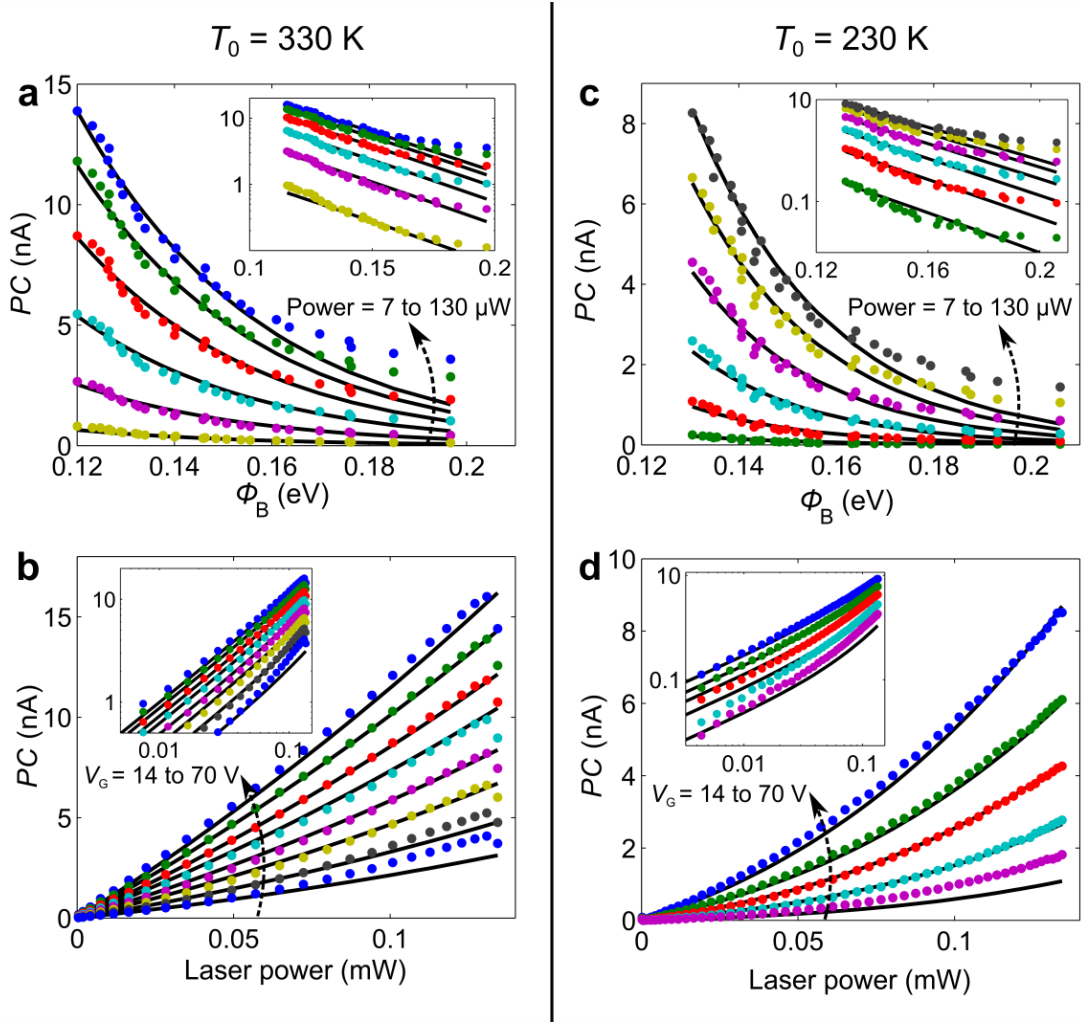

**Supplementary Figure 6: Comparison between the PTI model and the experimental photoresponse measured at $T_0$ = 230 and 330 K, with $\lambda$ = 1500 nm and $V_B$ = 0.36 V in a G/28-nm-thick WSe$_2$/G heterostructure. a,c)** *PC* vs. $\Phi_B$ at various laser powers and (**b,d**) *PC* vs. laser power *P* at different gate voltages $V_G$ measured at (**a,b**) $T_0$ = 330 K and (**c,d**) $T_0$ = 230 K. The data points correspond to the experiment and the solid lines to the model. Insets: Same experimental data and theoretical curves in logarithmic scale.

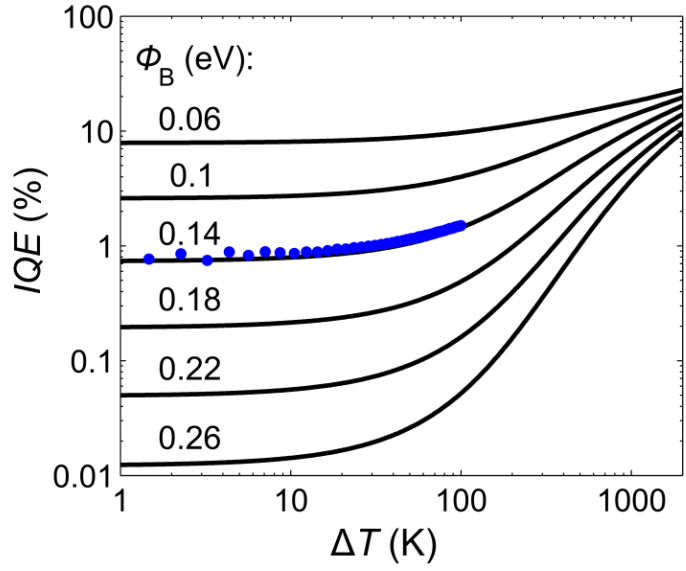

**Supplementary Figure 7: Internal Quantum Efficiency (*IQE*) vs. laser-induced temperature change *ΔT*.** The solid lines are theoretical *IQE* calculated with the PTI model (see Method section of main text) with $\tau_{inj}$ = 47 ps and $\Gamma$ = 0.5 MWm$^{-2}$K$^{-1}$, at $T_0$ = 300 K and at different values of $\Phi_B$ (indicated on top of each line). The data points are taken from the measurements presented in Figure 4b of the main text, at $\Phi$ = 0.14 eV.

# Supplementary Note 1: Cooling pathways of thermalized hot carriers in graphene

The heating and cooling of charge carriers in graphene is the subject of intense investigation and many different energy relaxation mechanisms have been suggested. For our experiment, we consider the energy pathways illustrated in Supplementary Figure 1a. Photons with an energy larger than twice the graphene Fermi level ($E_{photon} > 2\mu$) are absorbed in graphene due to interband transitions, creating photoexcited electrons and holes with energy $E = E_{photon}/2$. This energy is transferred to the phonon and electron bath typically through optical phonon emissions or carrier-carrier collisions, respectively. These processes are characterized by the energy-loss rate of the photoexcited carrier due to optical phonon emission $\Gamma_{e-op}$ and carrier-carrier scattering $\Gamma_{e-e}$ (Supplementary Figure 1a). Due to strong carrier-carrier interactions in graphene, carriers thermalize among themselves on ultrafast timescale (within 50 fs)[1], which leads to a branching ratio[2] between the two processes $\Gamma_{\text{e-e}}/\Gamma_{\text{e-op}}$ larger than 1. This implies that most of the absorbed photon energy is redistributed to the electron bath. Hence, in our experiment we consider that $\eta_{\text{heat}} = 70\%$ of the laser power absorbed in graphene is transferred to the electron bath[2], giving rise to a thermalized hot carrier distribution with temperature $T_{\text{e}}$.

Hot thermalized carriers subsequently cool down to equilibrate with the phonon (lattice) temperature $T_{\text{ph}}$ and the ambient temperature $T_0$. This cooling can be due to various processes such as the emission of intrinsic acoustic phonons[3], disorder-enhanced supercollisions with acoustic phonons[4,5], interaction with remote surface polar phonon modes (SPP) of the substrate[6] and in-plane heat dissipation via diffusion of hot carriers. All these mechanisms exhibit a different dependence on $T_{\text{e}}$. To simplify, we assume that for low increase in electronic temperature ($\Delta T = T_{\text{e}} - T_{\text{ph}} \ll T_{\text{ph}}$) the cooling of hot carriers is proportional to $\Delta T$ with an electron-phonon coupling constant $\Gamma_{\text{e-ph}}$. This coupling increases the phonon bath temperature $T_{\text{ph}}$ and is finally dissipated through the substrate (at temperature $T_0$) at a rate $\Gamma_0$. Hence, we model the temperatures of the electron and phonon bath using the following equations:

$$C_{\text{e}} \frac{\partial T_{\text{e}}}{\partial t} = P_{\text{in}}(t) - \Gamma_{\text{e-ph}}(T_{\text{e}} - T_{\text{ph}}) \qquad \text{(Supplementary Equation 1)}$$

$$C_{\text{ph}} \frac{\partial T_{\text{ph}}}{\partial t} = \Gamma_{\text{e-ph}}(T_{\text{e}} - T_{\text{ph}}) - \Gamma_0(T_{\text{ph}} - T_0) \qquad \text{(Supplementary Equation 2)}$$

where $C_{\text{e}}$ is the electronic heat capacity defined as the product of the 2D Sommerfeld constant and the electron temperature, $C_{\text{e}} = \gamma T_{\text{e}} = (2\pi\mu k_{\text{B}}^2/3\hbar^2 v_{\text{F}}^2)T_{\text{e}}$. Here, $k_{\text{B}}$ is the Boltzmann's constant, $v_{\text{F}}$ is graphene's Fermi velocity and $\hbar$ is the reduced Planck constant. $P_{\text{in}}(t)$ is the laser power density that is absorbed by graphene (approximately 0.5% absorption considering the dielectric permittivity of the surrounding medium[7]) and transferred to the electronic bath ($\eta_{\text{heat}} = 70\%$) (Supplementary ref. 2). $C_{\text{ph}}$ is the phonon heat capacity which we assume to be roughly $10^4$ times larger than $C_{\text{e}}$ (Supplementary ref. 8).

The temperature dynamics predicted by supplementary equations 1 and 2 largely depends on the rate-limiting relaxation step. Since the out-of-plane (c-axis) thermal conductivity of boron nitride[9] is approximately 2 Wm$^{-1}$K$^{-1}$, we estimate $\Gamma_0$ ~ 30 MWm$^{-2}$K$^{-1}$ for our 70-nm-thick hBN substrate. This value is much larger than theoretical estimates[10] of $\Gamma_{\text{e-ph}}$, which are typically between 0.5 and 5 MWm$^{-2}$K$^{-1}$. Therefore, the electron-phonon cooling creates a "bottleneck" that confines the heat in the electron bath. We also note that since the in-plane electronic thermal conductivity $\kappa_e$ is small ($\kappa_e$ ~ 1 Wm$^{-1}$K$^{-1}$), the cooling due to lateral diffusion of hot carriers (~$h\kappa_e/A$, where $h$ = 0.3 nm is the thickness of graphene and $A$ = 2.5 μm$^2$ is the laser spot size) is negligible (i.e., $\Gamma_{\text{e-ph}} \gg h\kappa_e/A$) (Supplementary ref. 10).

For steady-state conditions and $\Gamma_{\text{e-ph}} \ll \Gamma_0$, Supplementary Equations 1 and 2 simplify to
$$T_\text{e} - T_0 \cong P_\text{in}/\Gamma_{\text{e-ph}} \quad \text{(Supplementary Equation 3)}$$
We use this relation to model the PTI effect in the main text (see Methods) and find that the data is best described with $\Gamma_{\text{e-ph}}$ = 0.5 ± 0.3 MWm$^{-2}$K$^{-1}$. This value is compatible with the calculated out-of-plane thermal conductance of G/hBN interfaces caused by electron coupling with SPP phonons[6]. It is also consistent with a disorder-enhanced supercollision cooling mechanism[11] which predicts $\Gamma_{\text{e-ph}} \cong 3\Sigma T_{\text{ph}}^2$ under steady-state conditions[5], where $\Sigma$ is the supercollision rate coefficient (typically between 0.5 and 2 Wm$^{-2}$K$^{-3}$) (Supplementary ref. 4).

To verify the validity of Supplementary Equation 3 for our experimental conditions, we solve Supplementary Equations 1 and 2 to exactly calculate the variation in temperature of the electron ($\Delta T_\text{e}$ = $T_\text{e}$ - $T_0$) and phonon ($\Delta T_\text{ph}$ = $T_\text{ph}$ - $T_0$) baths induced by a quasi-CW laser pulse similar to the one used in our experiment (pulse duration $dt$ = 100 ps and repetition rate $f$ = 40 MHz) with average laser power $P$ = 100 μW, at $T_0$ =300 K and using $\Gamma_{\text{e-ph}}$ = 0.5 MWm$^{-2}$K$^{-1}$ and $\Gamma_0$ = 30 MWm$^{-2}$K$^{-1}$. The calculations (Supplementary Figure 1b) show that the $T_\text{e}$ reaches a temperature that is much higher than $T_\text{ph}$ (and $T_0$) and in quantitative agreement with Supplementary

Equation 3. Moreover, the temperature closely follows the variation of the pulse intensity in time. This confirms the validity of the steady-state approximation, which is expected since the measured cooling time $\tau_{\text{cool}}$ is much shorter than the pulse duration. More importantly, these calculations indicate that the electron bath is thermally decoupled from the phonon bath ($T_e > T_{\text{ph}}$) under (quasi-) steady-state conditions.

## Supplementary Note 2: Devices with different WSe₂ thicknesses

We have varied the thickness $L$ of the WSe$_2$ layer providing the energy barrier between the two graphene sheets. All the data shown in the main text come from a device containing WSe$_2$ layer with $L$ = 28 nm. We also measured sub-bandgap photocurrent on devices with $L$ = 2.2, 7.4 and 55 nm. These devices are made using the layer assembly technique described in Supplementary ref. 12, and are deposited on a Si/SiO$_2$ substrate that acts as a gate electrode. All devices display features in photocurrent that we attribute to the PTI effect. Supplementary Figures 2a and 3a show positive photocurrent autocorrelation peaks in devices with $L$ = 2.2 and 7.4 nm (for details on the measurement technique see Supplementary Note 4). As discussed in the main text, the dynamics of this peak is characterized by a time constant $\tau_{\text{cool}} \sim$ 1-2 ps which is on the order of the cooling time of the hot carriers in graphene. We also observe a superlinear power dependence of the photocurrent in these devices (Supplementary Figures 2c and 3c), which is characteristic of the PTI effect. Finally, the increase in photocurrent with gate voltage $V_G$ (Supplementary Figures 2d and 3d) is consistent with the PTI effect, which depends exponentially on the Schottky barrier height $\Phi_B$. This effect is also observed in the device with $L$ = 55 nm. Importantly, we note that the magnitude of the photocurrent does not vary significantly with the WSe$_2$ thickness $L$, indicating that tunneling effects (which depend exponentially on $L$) do not likely play a role in the photocurrent generation process. These observations further reinforce the conclusion that the PTI effect governs the photocurrent response of these heterostructures in this photon energy range.

## Supplementary Note 3: Internal photoemission in Au/WSe$_2$/Au heterostructures

In order to emphasize the difference between the internal photoemission (IPE) process typically observed at metal/semiconductor interface and the photo-thermionic (PTI) effect measured at graphene/WSe$_2$ junctions, we study the sub-bandgap photocurrent generated in Au/WSe$_2$/Au vertical heterostructures. Supplementary Figure 4a shows such a device made with a 20-nm-thick WSe$_2$ layer and a 10-nm-thick Au top electrode (similar measurements were obtained on a Au/40-nm-thick WSe$_2$/Au device). The photocurrent map in Supplementary Figure 4b shows that photocurrent is generated by sub-bandgap photons ($\lambda$ = 1500 nm) in the region where all three layers overlap. Interestingly, the magnitude of this photocurrent scales linearly with the laser power (Supplementary Figure 4c), regardless of the bias voltage $V_B$. This observation is consistent with the IPE process, which predicts that the number of carriers emitted over the barrier scales linearly with the number of initial photoexcited carriers. Moreover, Supplementary Figure 4d shows that the measured responsivity ($R = PC/P$) satisfies the relation $R \propto (E_{\text{photon}} - \phi_B)^2$ expected for the IPE process[13]. By projecting this relation to lower $E_{\text{photon}}$, we find a cut-off energy at $E_{\text{photon}} = \Phi_B \approx 0.4 - 0.5$ eV. We also note that the IPE responsivity of the Au/WSe$_2$/Au is smaller than the PTI responsivity measured in G/WSe$_2$/G, especially at low photon energy. For instance, at $E_{\text{photon}}$ = 0.8 eV, the maximum responsivity measured in Au-based device is $R$ = 0.036 mAW$^{-1}$, whereas graphene-based devices can easily reach $R$ = 0.12 mAW$^{-1}$.

All these observations that we attribute to IPE (i.e., linear power dependence, strong dependence on $E_{\text{photon}}$ and cut-off at $\Phi_B$) clearly contrast with the main features of the PTI effects (superlinear power dependence and no dependence on $E_{\text{photon}}$). This raises the question of why IPE dominates the photoresponse of metal/WSe$_2$ while PTI governs the one of graphene/WSe$_2$. The answer is two-fold. First, as discussed in the main text, carriers thermalize among themselves much more rapidly in graphene (approximately 10 fs) than in metal (approximately 100 fs) due to the stronger carrier-carrier interaction. This ultrafast thermalization process competes directly with the internal emission of initially photoexcited carriers (IPE). Assuming the timescale of this process to be equal for graphene and metal, we can conclude that IPE in graphene is suppressed by an order of magnitude compare to metal.

Secondly, we can explain why the PTI effect in graphene is larger than in metal from the result of the PTI model presented in the main text. Indeed, under steady-state conditions (e.i. when the pulse duration is much longer than $\tau_{cool}$, which is the case in our experiment except for time-resolved measurements), we can estimate the rise in electronic temperature as $\Delta T = P_{in}/\Gamma$, where $P_{in}$ is the incident power delivered to the carriers and $\Gamma$ is a thermal conductance term which describe the rate-limiting heat dissipation mechanism from the electronic system to the phonon bath (see Supplementary Note 1). From the fit of our PTI model we obtain $\Gamma_G = 0.5$ MWm$^{-2}$K$^{-1}$ for graphene/WSe$_2$, while the electron-phonon coupling constant of Au, for instance, is $\Gamma_{Au} = 2.6 \times 10^{16}$ Wm$^{-3}$K$^{-1}$ (Supplementary ref. 14). Considering the 10-nm-thick Au electrode of the device shown in Supplementary Figure 4a, this value corresponds roughly to a two-dimensional $\Gamma_{Au,2D}$ of 260 MWm$^{-2}$K$^{-1}$, which is two orders of magnitude larger than $\Gamma_G$. We conclude that for a given power $P_{in}$, the rise in electronic temperature $\Delta T$ is approximately 500 times larger in graphene than in gold (metals). This enhanced $\Delta T$, along with the suppression of the IPE due to ultrafast thermalization of carriers in graphene, explains why the PTI effect, which depends exponentially on the carrier temperature, dominates the photoresponse of graphene/semiconductors junctions but not the one of metal/semiconductors interfaces.

## Supplementary Note 4: Time-resolved photocurent measurements

Time-resolved photocurrent measurements are performed using a setup similar to the one described in Supplementary ref. 12 and shown in Supplementary Figure 5.

Pulses with a duration of approximately 200 fs and a spectral bandwidth of 200 nm centered at 800 nm are generated by a Ti:sapphire laser (Thorlabs Octavius) with a repetition rate of 85 MHz. Due to the broad emission spectrum of the laser, both photocurrent mechanisms (i.e. PTI due to graphene absorption and photoexcited charge transfer due to $WSe_2$ absorption) can contribute to the measured photocurrent autocorrelation signal. In order to isolate the PTI photocurrent and reduce the one originating from $WSe_2$ absorption, we insert an 800-nm long pass filter in the beam path. The filtered beam is split into two arms (one with a motorized delay stage) and recombined using 50/50 beam splitters. Time-resolved photocurrent measurements are performed by measuring the photocurrent (with a preamplifier and a lock-in amplifier synchronized with a mechanical chopper at 117 Hz) as a function of the time delay $\Delta t$ between the laser pulses of each arm.

We can further separate the contribution of the PTI signal (originating from graphene absorption) and the one of stemming from direct $WSe_2$ absorption by taking advantage of their different power dependence: the superlinear (sublinear) power dependence of graphene ($WSe_2$) absorption leads to a positive (negative) correlation signal[12]. Furthermore, under the right conditions (typically low temperature and bias voltage), the $WSe_2$ photocurrent can be suppressed and the signal indeed displays a positive peak around $\Delta t$ = 0 (see Figure 2d of main text, as well as Supplementary Figure 2a and 3a).

## Supplementary Note 5: PTI model and measurements at different ambient temperatures

To validate our model of the PTI effect, we repeat the measurement and analysis of PC vs. laser power $P$ and gate voltage $V_G$ at other ambient temperatures. At $T_0$ = 330 K (Supplementary Figures 6a and b), the fit of the model gives a carrier injection time $\tau_{inj}$ = 34 ± 10 ps and an out-of-plane interfacial thermal conductance $\Gamma$ = 1 ± 0.6 MWm$^{-2}$K$^{-1}$, whereas at $T_0$ = 230 K (Supplementary Figures 6c and d), we obtain $\tau_{inj}$ = 24 ± 10 ps and $\Gamma$ = 0.6 ± 0.3 MWm$^{-2}$K$^{-1}$. We find a good agreement between the model and experiment for both temperatures, but a more detailed study is needed in order to examine and understand the possible dependence of $\tau_{\text{inj}}$ and $\Gamma$ on temperature $T_0$.

## Supplementary Note 6: Estimation of the Internal Quantum Efficiency

Using the PTI model described in the main text, we can estimate the performance of the G/WSe$_2$/G heterostructure at higher electronic temperatures and lower Schottky barriers. Supplementary Figure 7 shows the predicted internal quantum efficiency (*IQE*, defined as the ratio between the number of carriers collected and the number of photons absorbed by the graphene layer) as a function of the change in temperature $\Delta T = T_e - T_0$ induced by our supercontinuum laser (see Method of the main text) at $\lambda$ = 1500 nm. Assuming ~0.5% absorption[7] we find that, in those conditions, the *IQE* can reach up to about 20% for small Schottky barriers $\Phi_B$ (smaller than 0.14 eV). We note however that our model does not take into account a possible change in thermal conductance $\Gamma$ with increasing $T_e$ and is only valid for $k_B T_e \ll \Phi_B$.

## Supplementary Note 7: Landauer transport model

To model the injection of hot carriers over the G/WSe$_2$ Schottky barrier, we consider the system shown in Figure 1b of the main text, which consists of a single G$_B$/WSe$_2$ heterojunction. We note that this system does not include the top graphene layer. Indeed, as explained in the main text, the contribution of the top graphene layer can be neglected when a positive interlayer bias $V_B$ is applied. In these conditions, according to Landauer's transport theory[15,16], the current density $J$ flowing through the heterojunction (considered as the resistor channel) between the graphene and WSe$_2$ layers can be written as:

$$J = \frac{e_0}{\tau_{\text{inj}}} \int_{-\infty}^{\infty} T(E) D(E) \big(f_G(E) - f_{\text{WSe}_2}(E)\big) dE , \quad \text{(Supplementary Equation 4)}$$

where $e_0$ is the elementary charge, $\tau_{inj}$ is the time it takes for an electron to transfer through the junction, referred here as the charge injection time, $T(E)$ is the transmission probability, and $f_G(E)$ and $f_{WSe_2}(E)$ are the Fermi-Dirac function of graphene and WSe$_2$, respectively. $D(E)$ is the density of states of graphene, where $\hbar$ is the reduced Planck's constant and $v_F$ is the graphene Fermi velocity.

We set the charge neutrality point of graphene to $E = 0$ (as illustrated in Figure 1b of the main text) and assume the tunneling contribution to the photocurrent to be negligible (which is the case for the thick WSe$_2$ layer considered here) and a unity transmission for energies above the Schottky barrier. We obtain:

$D(E) = 2|E|/\pi(\hbar v_F)^2$

$f_G(E) = 1/(e^{(E-\mu)/k_B T}+1)$

$f_{WSe_2}(E) = 1/(e^{(E-\mu+e_0 V_B)/k_B T}+1)$ (Supplementary Equation 5)

and

$T(E) = \begin{cases} 1 \text{ for } E > \phi_0 \\ 0 \text{ for } E \leq \phi_0 \end{cases}$

where $\mu$ is the Fermi energy of graphene, $V_B$ is the voltage applied across the WSe$_2$ layer (see Figure 1c of the main text) and $\phi_0$ is the offset between the WSe$_2$ conduction edge and graphene's Dirac point (see Figure 1b of the main text). In reverse bias condition (such that $f_G(E) \gg f_{WSe_2}(E)$ for $E > \phi_0$), Supplementary Equation 4 becomes:

$J = \frac{2}{\pi} \frac{e_0}{\tau_{inj}} \frac{1}{(\hbar v_F)^2} \int_{\phi_0}^{\infty} |E| f_G(E) dE$ (Supplementary Equation 6)

Supplementary Equation (6) can be solved analytically provided that $\phi_B = \phi_0 - \mu \gg k_B T$ and we obtain the equation of the PTI emission model presented in the Methods section of the main text:

$J = \frac{2}{\pi} \frac{e_0}{\tau_{inj}} \left(\frac{k_B T}{\hbar v_F}\right)^2 \left(\frac{\phi_0}{k_B T}+1\right) exp\left(\frac{-\phi_B}{k_B T}\right)$ (Supplementary Equation 7)

## Supplementary References

1. Tielrooij, K. J. *et al.* Generation of photovoltage in graphene on a femtosecond timescale through efficient carrier heating. *Nature Nanotech.* **10,** 437–443 (2015).

2. Tielrooij, K. J. *et al.* Photoexcitation cascade and multiple hot-carrier generation in graphene. *Nature Phys.* **9,** 248–252 (2013).

3. Bistritzer, R. & MacDonald, A. H. Electronic cooling in graphene. *Phys. Rev. Lett.* **102,** 206410 (2009).

4. Betz, A. C. *et al.* Supercollision cooling in undoped graphene. *Nature Phys.* **9,** 109–112 (2012).

5. Graham, M. W., Shi, S.-F., Ralph, D. C., Park, J. & McEuen, P. L. Photocurrent measurements of supercollision cooling in graphene. *Nature Phys.* **9,** 103–108 (2012).

6. Low, T., Perebeinos, V., Kim, R., Freitag, M. & Avouris, P. Cooling of photoexcited carriers in graphene by internal and substrate phonons. *Phys. Rev. B* **86,** 045413 (2012).

7. Stauber, T., Peres, N. M. R. & Geim, A. K. Optical conductivity of graphene in the visible region of the spectrum. *Phys. Rev. B* **78,** 085432 (2008).

8. Benedict, L. X., Louie, S. G. & Cohen, M. L. Heat capacity of carbon nanotubes. *Solid State Commun.* **100,** 177–180 (1996).

9. Simpson, A. & Stuckes, A. D. The Thermal Conductivity of Highly Oriented Pyrolytic Boron Nitride. *J. Phys. C Solid State Phys.* **4,** 1710–1718 (1971).

10. Freitag, M., Low, T. & Avouris, P. Increased responsivity of suspended graphene photodetectors. *Nano Lett.* **13,** 1644–1648 (2013).

11. Song, J. C. W., Reizer, M. Y. & Levitov, L. S. Disorder-Assisted Electron-Phonon Scattering and Cooling Pathways in Graphene. *Phys. Rev. Lett.* **109,** 106602 (2012).

12. Massicotte, M. *et al.* Picosecond photoresponse in van der Waals heterostructures. *Nature Nanotech.* **11,** 42–46 (2015).

13. Sze, S. M., Crowell, C. R. & Kahng, D. Photoelectric determination of the image force dielectric constant for hot electrons in Schottky barriers. *J. Appl. Phys.* **35,** 2534–2536 (1964).

14. Qiu, T. Q. & Tien, C. L. Heat Transfer Mechanisms During Short-Pulse Laser Heating of Metals. *J. Heat Transfer* **115,** 835-841 (1993).

15. Datta, S. Lessons from Nanoelectronics: a New Perspective on Transport (Wold Scientific, 2012).

16. Sinha, D. & Lee, J. U. Ideal Graphene/Silicon Schottky Junction Diodes. *Nano Lett.* **14,** 4660–4664 (2014).